\documentclass[showpacs,preprintnumbers,prl,twocolumn]{revtex4}
\usepackage{amssymb}
\usepackage{amsmath}
\usepackage{graphicx}
\usepackage{bm}
\usepackage{amsfonts}

\newcommand{\ket}[1]{\mbox{$|#1\rangle$}}

\def\be{\begin{equation}}
\def\ee{\end{equation}}

\begin{document}

\title
{State tomography of capacitively shunted phase qubits with high fidelity}

\author{Matthias Steffen, M. Ansmann, R. McDermott, N. Katz, Radoslaw C. Bialczak, Erik Lucero, Matthew Neeley, E.M. Weig, A.N. Cleland, John M. Martinis}

\email{martinis@physics.ucsb.edu}

\affiliation
{Department of Physics and California Nanosystems Institute, University of California, Santa Barbara, CA 93106, USA}



\begin{abstract}
We introduce a new design concept for superconducting quantum bits
(qubits) in which we explicitly separate the capacitive element from
the Josephson tunnel junction for improved qubit performance. The
number of two-level systems (TLS) that couple to the qubit is
thereby reduced by an order of magnitude and the measurement
fidelity improves to $90\%$. This improved design enables the first
demonstration of quantum state tomography with superconducting
qubits using single shot measurements.
\end{abstract}

\volumeyear{year}
\volumenumber{number} \issuenumber{number}
\eid{identifier}
\date{\today}

\maketitle

Superconducting circuits containing Josephson junctions provide a
promising approach towards the construction of a scalable solid-state
quantum computer \cite{Nakamura03_cnot, Vion02, Collin04, Chiorescu03,
Bertet05, Schoelkopf05}. The phase qubit \cite{Martinis02} has
significant potential because coupled qubits have been measured
simultaneously \cite{McDermott05} and the coherence times are
reasonably long \cite{Martinis05}. The conventional design of phase
qubits relies on the Josephson inductance and the self-capacitance of
the Josephson tunnel junction to form a non-linear microwave resonator
\cite{Martinis02}. Losses and noise in either component compromise
qubit performance, however, generally the inductive element has been
suspected as the root source of decoherence \cite{Vanharlingen04,
Wellstood04, Mueck05}.  It therefore came as a surprise when the
capacitive element was clearly identified as a main source of
decoherence \cite{Martinis05}. Because of the large intrinsic loss of
the junction capacitor, we believe it is not possible to fabricate
high fidelity phase qubits using a standard design and amorphous
aluminum oxide (AlO$_x$) tunnel barriers.

Can we redesign phase qubits to circumvent or minimize capacitive
losses without adversely affecting other desirable qubit properties?
Understanding this question has profound implications for other
designs of superconducting qubits, and may shed light on methods for
reducing noise from two-levels systems (TLS).

\begin{figure}[t]
\vspace*{-0.125in}
\begin{center}
\mbox{\includegraphics*[width=3.0in]{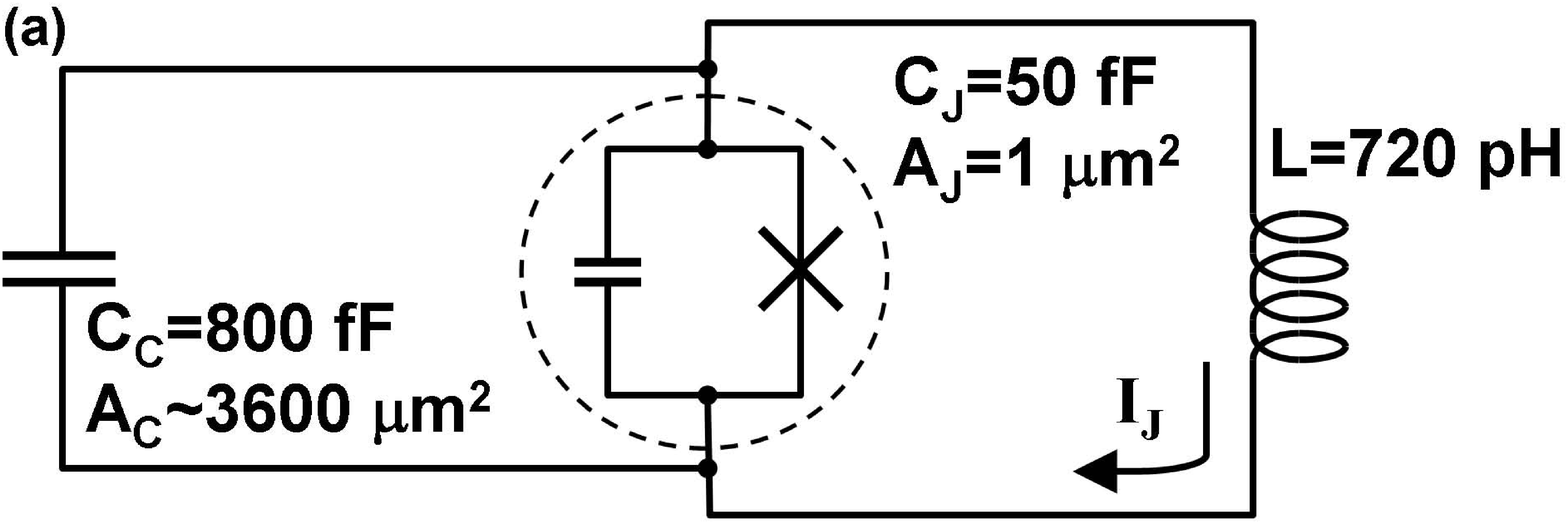}} \\
$ \noindent \begin{array}{ll}
\noindent \mbox{\includegraphics*[width=1.724in]{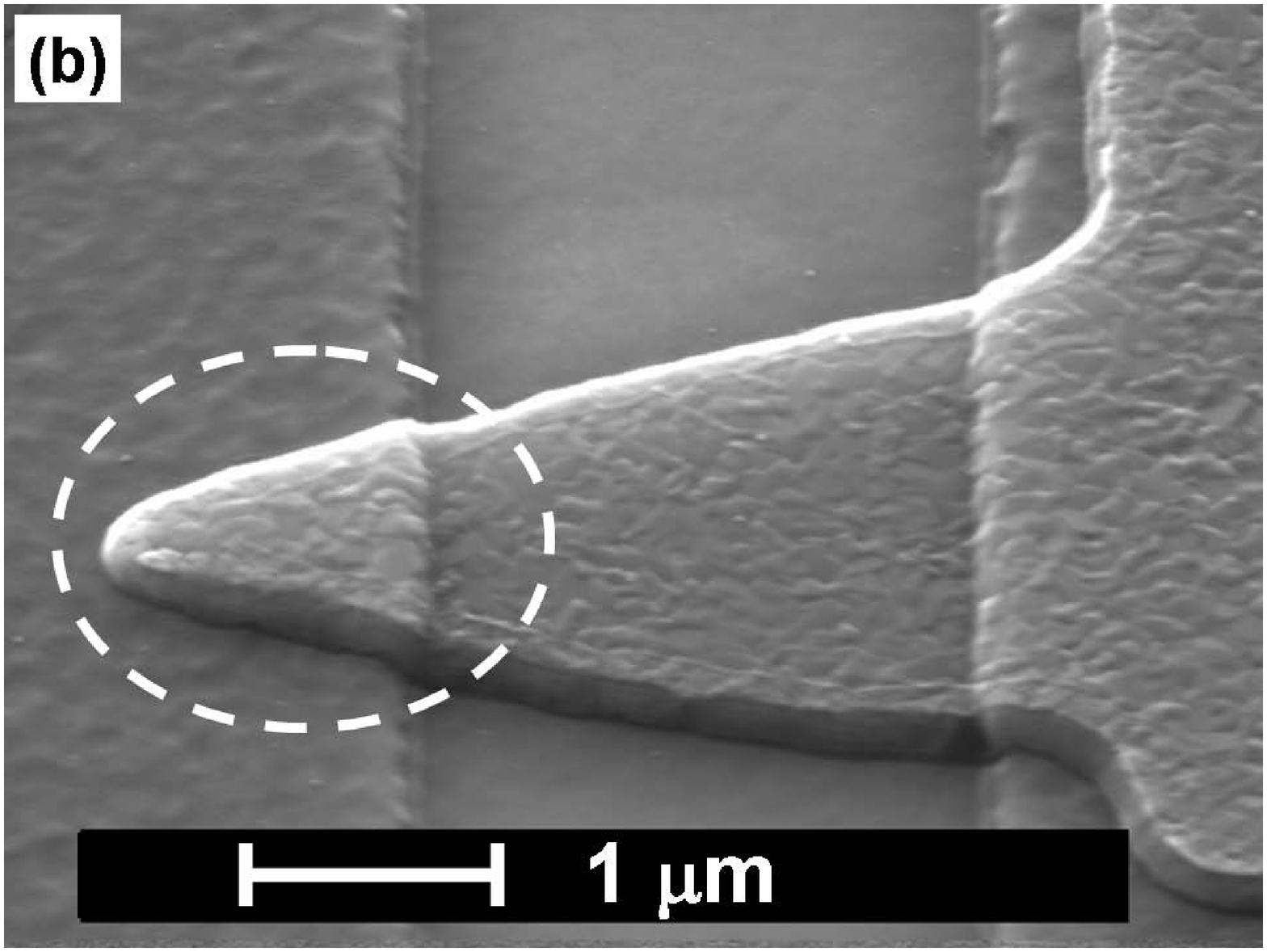}} &
\mbox{\includegraphics*[width=1.2in]{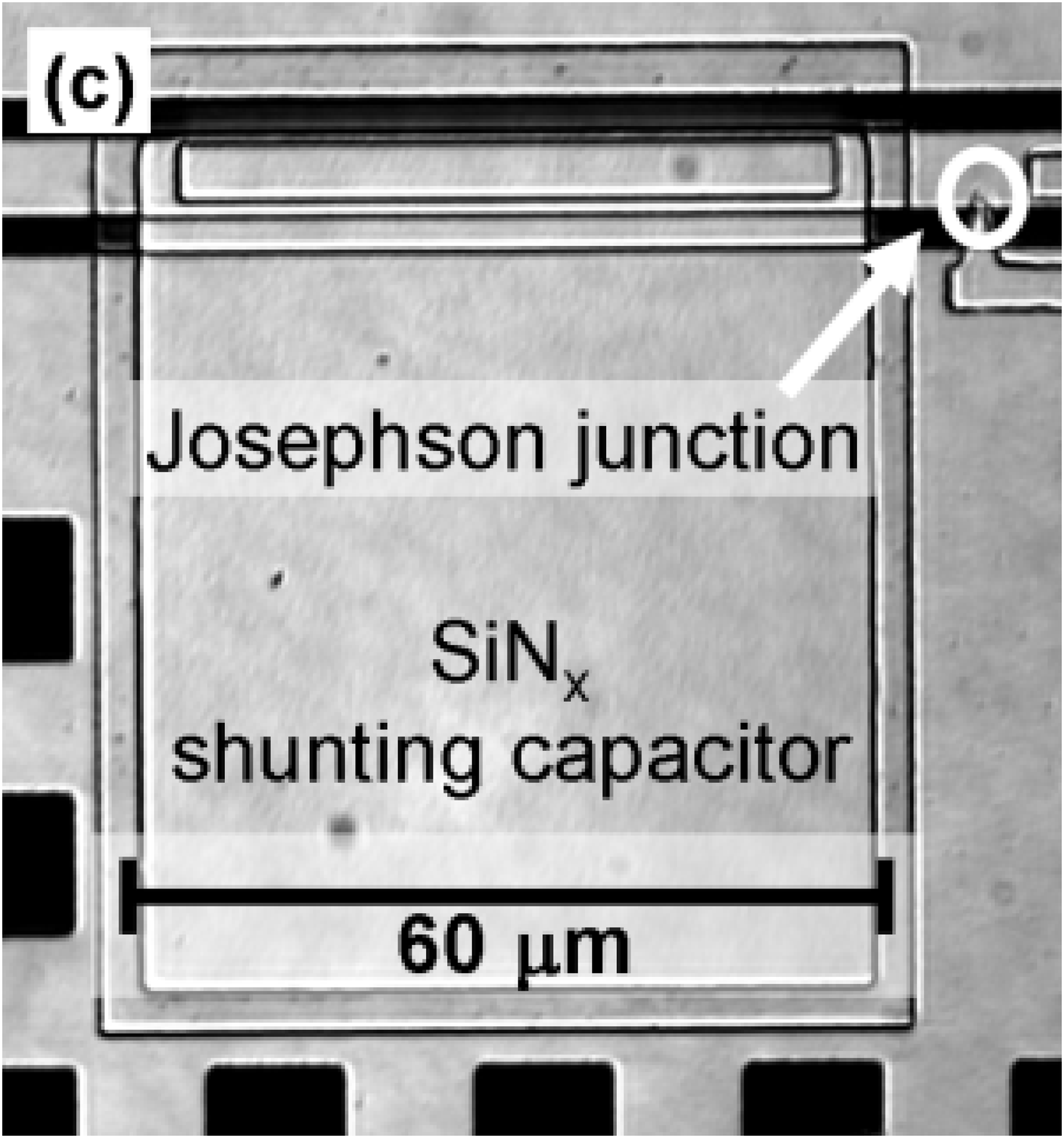}}
\end{array} $
\end{center}
\vspace*{-0.25in} \caption{Circuit diagram and micrograph of the
redesigned phase qubit. (a) A small area ($A_J \sim 1\
\mu\textrm{m}^2$) Josephson junction with little self-capacitance
($C_J=50\ \textrm{fF}$) is shunted by a large ($A_C \sim 3600\
\mu\textrm{m}^2$) high quality capacitor with $C_C=800\ \textrm{fF}$.
The qubit bias current $I_{J}$ is induced through an inductor with
$L=720\ \textrm{pH}$. (b) A scanning electron microscope image of
the small area Josephson junction shows well defined features at
sub-micron length scales. (c) An optical image of the shunting
capacitor and the Josephson junction (white circle).}
\label{fig:circuit}
\end{figure}

Here, we introduce a new design concept for superconducting phase
qubits with which further progress can be achieved in a
straightforward manner. The central idea is to explicitly separate
out the capacitive from the inductive element of the Josephson
junction, allowing their properties to be separately optimized. This
idea can be realized by shunting a tunnel junction, which has little
self-capacitance but the same Josephson inductance as the
conventional design, with a capacitor which has a lower intrinsic
loss tangent than the original junction capacitor, as sketched in
Fig.~\ref{fig:circuit}(a-c). We have fabricated an improved
generation of phase qubits whose measurement fidelity is
significantly improved. This success enables the first demonstration
of quantum state tomography using a superconducting qubit with
single shot measurements.

\begin{figure}[t]
\vspace*{-0.125in}
\begin{center}
\mbox{\includegraphics*[width=3.0in]{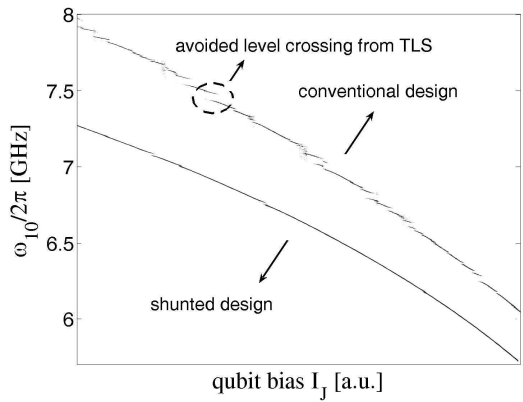}}
\end{center}
\vspace*{-0.25in} \caption{Qubit frequency $\omega_{10}/2\pi$ versus
qubit bias for the conventional single-element design (top trace,
shifted for clarity) and the capacitively shunted design (bottom
trace). The bias current $I_{J}$ tunes the frequency of the qubit
\cite{Simmonds04,Cooper04}. The density of splittings in the
conventional design is high enough that the qubit is almost always
coupled to at least one TLS. The state occupation of $\ket{1}$ is
consequently diminished, giving rise to decoherence and thus
reducing qubit performance. The new design has about one tenth of
the area and consequently exhibits a reduction in the number of
avoided level crossings by approximately a factor of ten.}
\label{fig:spect}
\end{figure}

Progress using phase qubits has thus far been hindered by a large
density of TLS defects that couple to the qubit\cite{Martinis05}.
Individual TLS manifest themselves as avoided level crossings
(splittings) in the qubit spectroscopy, as shown in Fig.
\ref{fig:spect}. The defects are located in the insulating barrier of
the tunnel junction, and have a significant density because of the
relatively large intrinsic loss tangent $\delta_i \sim 1.6 \cdot
10^{-3}$ of AlO$_x$.  Our model predicts that high fidelity phase
qubits are not possible using a standard design and amorphous AlO$_x$
tunnel barriers.

However, the number of TLS defects may be dramatically reduced with
redesign.  Loss of coherence from TLS depends both on the density and
size of the splittings, and for low density scales as the square of
the tunnel-junction area divided by the total
capacitance\cite{tlsnote,Martinis05}. A dramatic improvement in
fidelity can thus be achieved by reducing the area of the tunnel
junction from $\sim 10\ \mu \mathrm{m}^2$ to $\sim 1\ \mu
\mathrm{m}^2$ while holding its critical-current constant, and keeping
the total capacitance constant by adding an external low-loss
capacitor. With a decrease in the density of splittings, measurement
errors \cite{Cooper04} are also predicted to lower by a factor of 10.
The resulting qubit design of Fig. \ref{fig:circuit} therefore simply
substitutes the lossy capacitance from the tunnel junction with an
external element of higher quality. Note that for the standard (single
element) design, the junction area scales with the total capacitance
and thus loss in coherence scales just as the junction area. The
ability to separately control total capacitance and junction area in
the new design therefore allows improved protection from decoherence
due to individual TLS, in addition to increasing measurement fidelity.

\begin{figure}[t]
\vspace*{-0.125in}
\begin{center}
\mbox{\includegraphics*[width=3.5in]{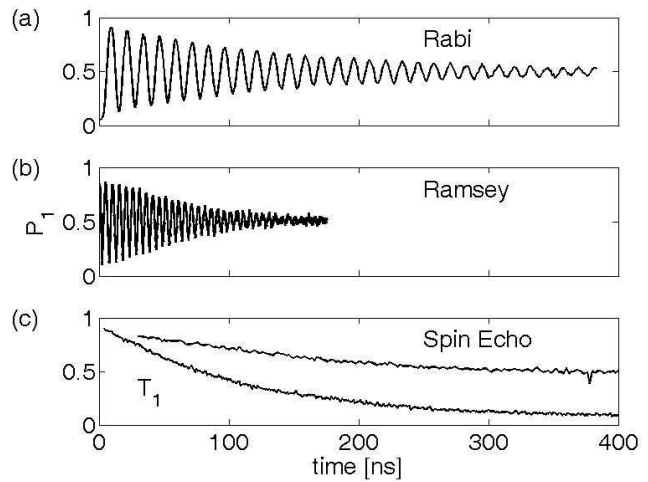}}
\end{center}
\vspace*{-0.25in} \caption{Experimental results using the capacitively
shunted phase qubit design. The vertical axis of each plot displays
the probability $P_1$ of occupying the $\ket{1}$ state. (a) Rabi
oscillations have a visibility of about $85 \%$ and decay at a rate
consistent with the measured qubit relaxation time of $T_1 \approx
110\ \textrm{ns}$ (shown in (c)). (b) The Ramsey fringes decay
non-exponentially and are consistent with spectroscopic linewidths. We
extract a dephasing time of $T_2^* \approx 90\ \textrm{ns}$. (c) The
impact of low-frequency 1/f noise can be significantly reduced using a
Spin Echo sequence. We extract an intrinsic dephasing time of $T_2
\approx 160\ \textrm{ns}$. Measurement of the energy decay ($T_1$) is
also plotted.}
\label{fig:rabramspin}
\end{figure}

The redesign requires reliable fabrication of tunnel junctions with an
area of about $\sim 1\ \mu\textrm{m}^2$. The tunnel barrier is formed
by an Ar ion-mill clean followed by thermal oxidation of the Al base
electrode. The Josephson junction is next defined by optical
lithography and a reactive ion etch in an Ar/Cl plasma. A micrograph
of the small-area junction is shown in Fig. \ref{fig:circuit}(b).  The
shunting capacitor, shown in Fig. \ref{fig:circuit}(c), is made of
silicon nitride.  It was previously shown to have twenty times lower
dielectric loss than SiO$_2$ \cite{Martinis05} and thus is an
acceptable choice for this first demonstration.

Experimental tests \cite{Cooper04,Simmonds04} on the redesigned phase
qubit confirm the expected behavior. The number of splittings visible
in the qubit spectroscopy is reduced roughly by an order of magnitude,
yet their sizes are comparable to those of the conventional design
(see Fig. \ref{fig:spect}). The visibility of the Rabi oscillations
(Fig. \ref{fig:rabramspin}(a)) is about $85 \%$ and the decay is
limited by the measured energy relaxation time of $T_1 \approx 110\
\textrm{ns}$. Because of the small number of splittings, we are able
to increase the Rabi oscillation period to about $12.5\ \textrm{ns}$
(limited by pulse shaping and the detuning of $\omega_{21}$
\cite{Steffen03}), compared to a $\sim 50\ \textrm{ns}$ Rabi
oscillation period using the conventional design \cite{Martinis05}.

From the data, we compute a measurement fidelity that is close to
$90 \%$. This is an improvement of about $40 \%$ compared with the
measurement fidelity using the standard phase qubit design. Our
observations confirm the prediction that measurement fidelity is
reduced by sweeping through avoided level crossings during our
measurement pulse \cite{Cooper04}. Half of the remaining $10 \%$
loss in the measurement fidelity can be attributed to energy
relaxation (as the measurement takes about $5\ \textrm{ns}$), while
the remaining half can be attributed to a loss in signal due to
sweeping through a few remaining junction resonances. Additional
experimental data, including Ramsey fringes and a Spin Echo (Fig.
\ref{fig:rabramspin}(b) and (c)) are consistent with measured
linewidths from the qubit spectroscopy as well as the energy
relaxation time.

The energy relaxation of the qubit is limited by the SiN$_x$
shunting capacitor, which has a measured loss tangent of $\delta_i
\approx 1.5 \cdot 10^{-4}$. With a qubit frequency $\omega_{10}/2\pi
\approx 6$ GHz, we expect an energy relaxation of $T_1 = 1/\delta_i
\omega_{10} \approx 170\ \textrm{ns}$, which is close to the
measured value. Further improvements in $T_1$ are possible by
fabricating shunting capacitors with even lower loss tangents.

The demonstrated improvements in fidelity now enable state tomography,
which is necessary for a full characterization of qubit states and
gate operations. Typically, state tomography involves measuring an
unknown quantum state in different basis sets to extract its location
on the Bloch sphere. However, in our experiment it is more convenient
to always measure in the $\ket{0}$ and $\ket{1}$ basis, and perform
the basis change through single qubit rotations prior to measurement.

\begin{figure}[t]
\vspace*{-0.125in}
\begin{center}
\mbox{\includegraphics*[width=3.5in]{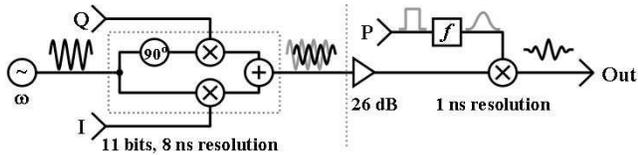}}
\end{center}
\vspace*{-0.25in} \caption{Schematic diagram of the microwave
pulser. The amplitudes of the two quadrature components $I$ and $Q$
are added to obtain microwaves with adjustable amplitude and phase,
and are then amplified by 26 dB. A filtered digital pulse is then used
in a second mixer to produce a Gaussian-shaped microwave pulse.}
\label{fig:iqmixer}
\end{figure}

To perform tomography, we have designed and built a custom microwave
pulse sequencer capable of producing microwave pulses with arbitrary
amplitude and phase (see Fig. \ref{fig:iqmixer}). We use an IQ mixer
that adds, with separate amplitude control, the two quadrature
components $I$ and $Q$ of a continuous-wave microwave signal. The
amplitude of both phases are controlled by an 11-bit digital to
analog (DAC) converter that may be updated every $8\ \textrm{ns}$. A
second mixer performs gating and pulse shaping with a time
resolution of $1\ \textrm{ns}$.

We use this microwave pulse sequencer to implement state tomography
with two different techniques. The first method simply rotates an
unknown Bloch vector over all rotation amplitudes and angles in the
$x-y$ plane of the Bloch sphere prior to measuring the final state
occupation probability $P_1$ of $\ket{1}$. From the resulting
two-dimensional probability map, the direction of the Bloch vector can
be simply computed from the amplitude and phase of the DAC values at
maximum $P_1$, and the Bloch vector length from the maximum contrast
of $P_1$.

\begin{figure}[t]
\vspace*{-0.125in}
\begin{center}
\mbox{\includegraphics*[width=3.4in]{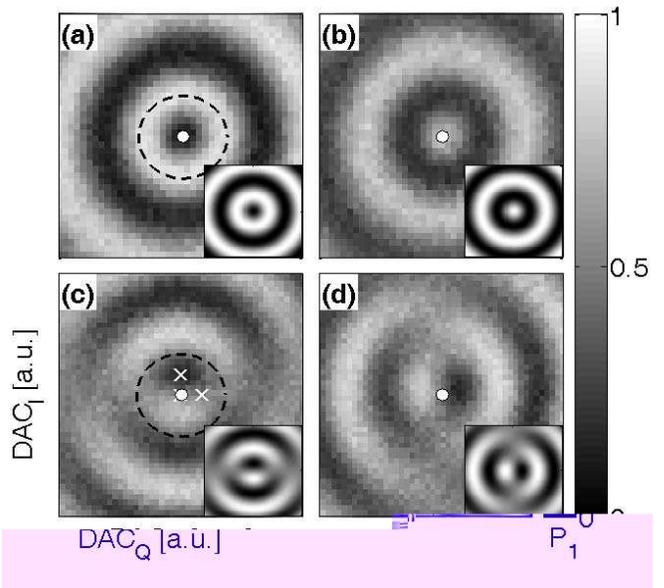}}
\end{center}
\vspace*{-0.25in} \caption{Probability $P_1$ of occupying state
$\ket{1}$ versus DAC$_I$ and DAC$_Q$, which control the amplitude of
the $\hat y$ and $\hat x$ tomography rotations, for the input states
(a) \ket{0}, (b) $\ket{1}$, (c) $(\ket{0}+\ket{1})/\sqrt{2}$, and
(d) $(\ket{0} +i\ket{1})/\sqrt{2}$. The dashed black circle
indicates a total pulse amplitude corresponding to a $\pi$-pulse at
different phases. The white dots label the zero amplitude points,
and the three white crosses indicate the measurements necessary for
a simplified state tomography technique. The inset in each plot
shows the predicted probability map.} \label{fig:tomo}
\end{figure}

In Fig. \ref{fig:tomo} we plot $P_1$ as a function of amplitude of
the $I$ and $Q$ components for several different initial states. In
Fig. \ref{fig:tomo}(a) the initial state is the ground state
$\ket{0}$, and we observe generalized Rabi oscillations that are
independent of the microwave phase, as expected. For the $\ket{1}$
state of Fig. \ref{fig:tomo}(b), the populations are inverted as
compared to (a). In Fig. \ref{fig:tomo}(c) the initial state is the
superposition $(\ket{0} + \ket{1})/\sqrt 2$, which is pointing along
the $\hat x$-direction on the Bloch sphere. As a result, rotations
about the $\hat x$-direction (controlled by DAC$_Q$) will not affect
the state. However, rotations along the $\hat y$-direction
(controlled by DAC$_I$) rotate the state and result in Rabi
oscillations. The state $(\ket{0} + i\ket{1})/\sqrt 2$ is similarly
plotted in (d), and shows a $90^\circ$ rotation in the direction of
oscillations as compared to (c). Theoretical predictions are shown
in the insets of Fig. \ref{fig:tomo}; small differences arise
primarily from off-resonance effects of the $\ket{2}$ state
\cite{Steffen03}.

The data clearly shows this technique can be used to reconstruct an
unknown quantum state for \textit{complete} testing of one qubit.
However, a more efficient and widely used technique makes use of the
fact that any arbitrary single qubit density matrix can be described
by $\rho = (I + \sum_{i=1}^3 c_i \sigma_i)/2$ where $\sigma_i$ denotes
the Pauli matrices, $I$ is the identity matrix, and $c_i$ are real
coefficients \cite{Chuang98,Liu05}. Therefore, in order to reconstruct
an unknown quantum state, we must identify the three-component vector
$\overrightarrow{c}$.

The simplest approach measures the three components $c_i$ directly by
performing three different read-out schemes. The first read-out is a
simple measurement of the state occupation probability of $\ket{1}$,
which measures the $\sigma_z$ component. The second (third) read-out
applies a $90^\circ$ rotation around the $\hat x$-direction ($\hat
y$-direction), followed by a measurement of the occupation probability
of $\ket{1}$, measuring the $\sigma_y$ ($\sigma_x$) components. Using
the three measured occupation probabilities we can reconstruct the
quantum state via a least squares fit and place it on the Bloch
sphere. Experimentally, this technique is advantageous because it
requires only three different read-out sequences instead of acquiring
a two-dimensional map.

\begin{figure}[t]
\vspace*{-0.125in}
\begin{center}
\mbox{\includegraphics*[width=3.4in]{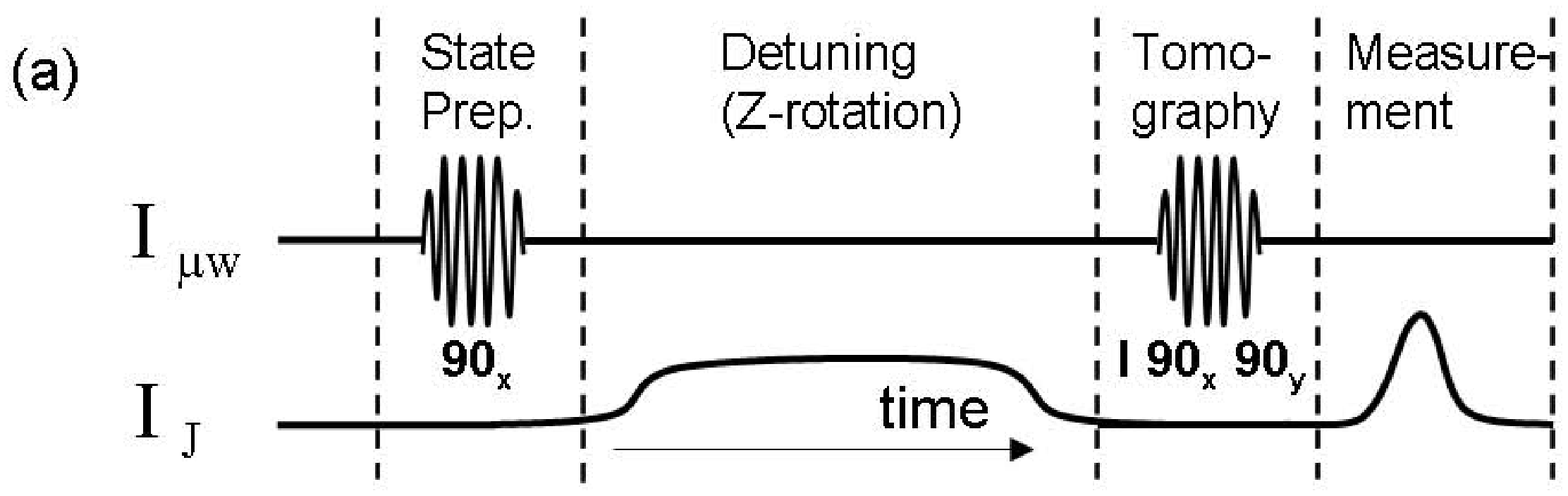}}
\mbox{\includegraphics*[width=3.4in]{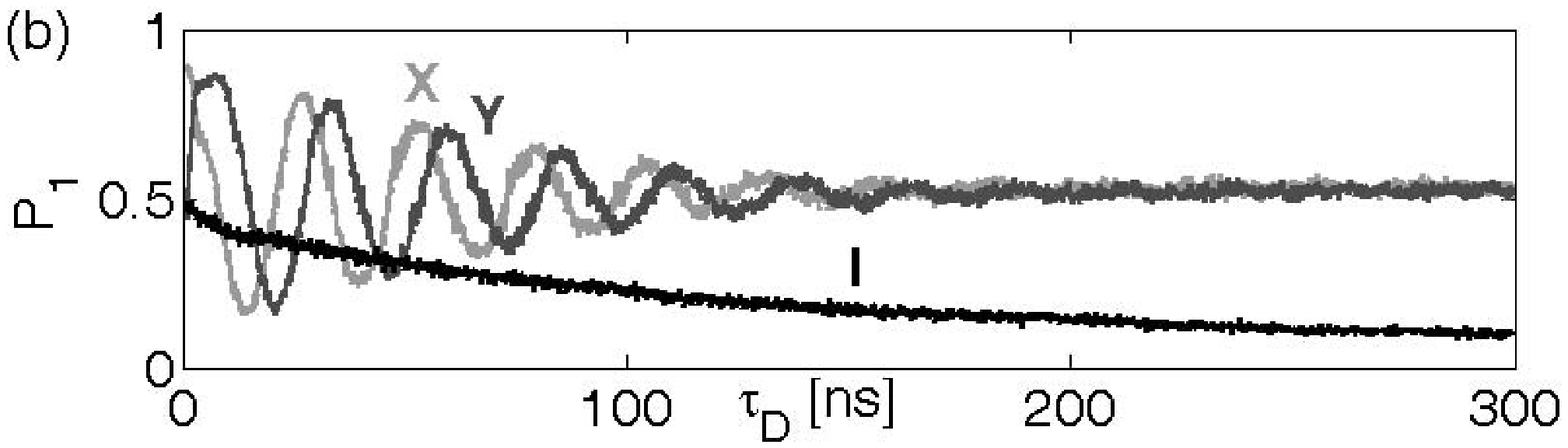}}
\end{center}
\vspace*{-0.55in} 
\begin{center}
\mbox{\includegraphics*[width=3.4in]{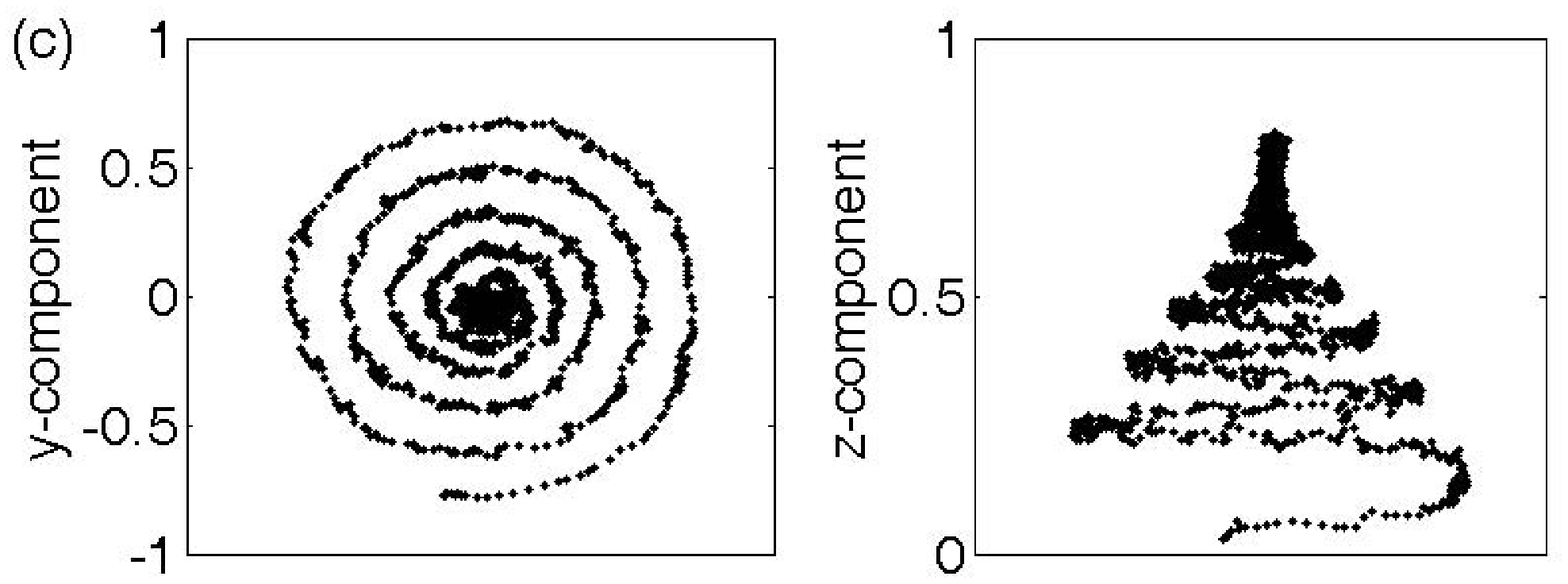}}
\end{center}
\vspace*{-0.5in} 
\begin{center}
\mbox{\includegraphics*[width=3.4in]{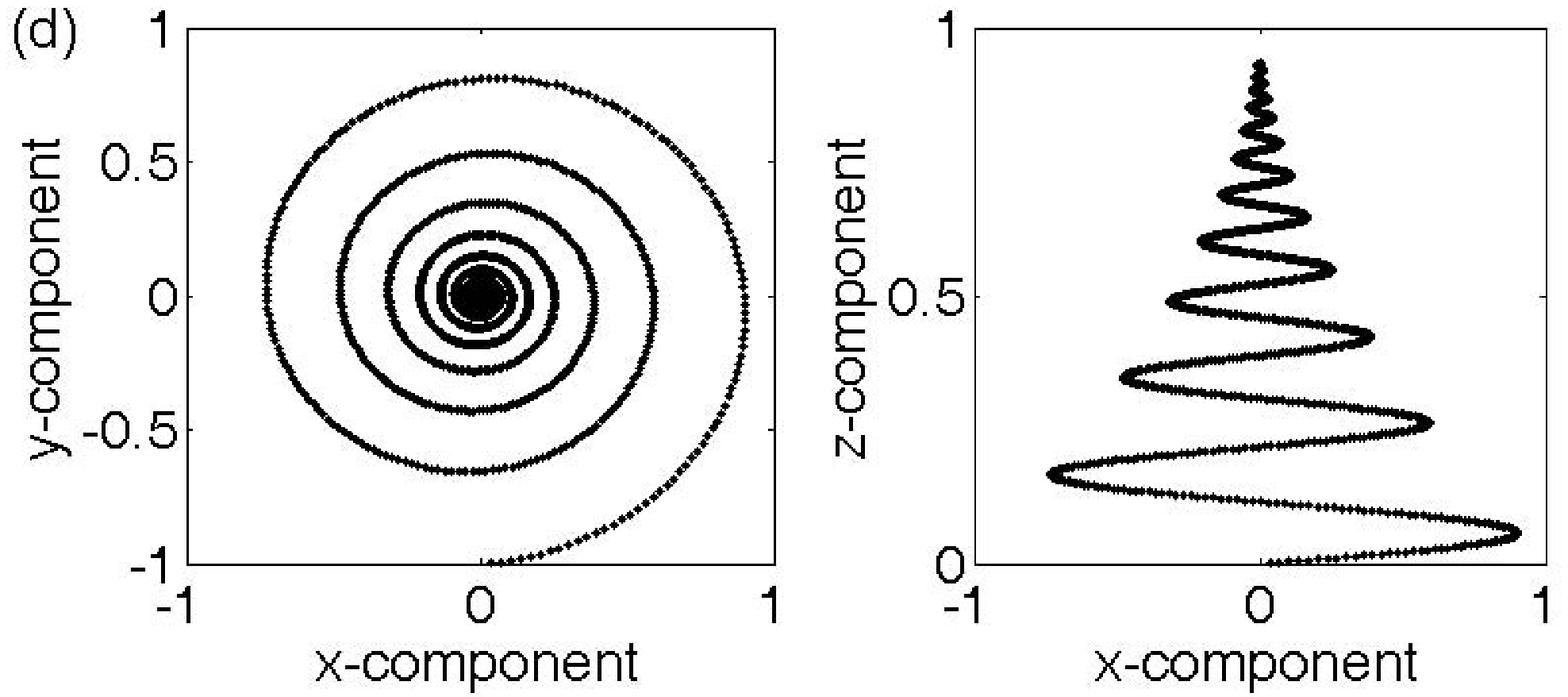}}
\end{center}
\vspace*{-0.25in} \caption{State tomography of the Bloch vector during
a Ramsey-type experiment. (a) Sequence of operations of the
experiment. A microwave pulse $I_{\mu w}$ at frequency
$\omega_{10}/2\pi=5.838$ GHz is first applied to give a $90^\circ$
rotation about the $\hat x$-axis.  An adiabatic bias pulse $I_{J}$
\cite{Nadav06} is then applied for a time $\tau_d$, during which the
qubit is detuned by $37$ MHz. The final $90^\circ$ tomography pulses
are $4\ \textrm{ns}$ long, which are short compared with the
relaxation times. Each data point was taken with 2,000 statistics and
is separated in time by $0.1\ \textrm{ns}$. (b) Plot of $P_1$ versus
$\tau_d$ for the three tomography pulses $90_X$, $90_Y$, and $I$. (c)
Projections of the reconstructed quantum state on the xy and xz-planes
of the Bloch sphere. (d) Theoretical prediction of the evolution of
the Bloch vector including relaxation using $T_1=110\ \textrm{ns}$ and
$T_2^*=90\ \textrm{ns}$.} 
\label{fig:ramsey}
\end{figure}

We use this state tomography to trace out the time evolution of the
single qubit quantum state in a Ramsey type experiment, as shown in
Fig. \ref{fig:ramsey}. In this sequence, we include a current pulse
during free evolution to rotate the qubit state about the $\hat
z$-axis.  This rotation is observed experimentally, and dephasing
causes the $xy$-component to shrink (Fig. \ref{fig:ramsey}c, left
panel) and relax toward the ground state (Fig. \ref{fig:ramsey}c,
right panel). These effects are confirmed by a theoretical model
which includes energy relaxation and dephasing (see Fig.
\ref{fig:ramsey}d). The main difference between the experiment and
theory is that the length of the reconstructed vector of the first
point is about $80 \%$ instead of unity. The loss is explained by
the $90 \%$ measurement fidelity and a time delay of $8\
\textrm{ns}$ between the end of the detuning and the tomography
pulses resulting in an additional $\sim 10 \%$ loss. The loss due to
reduced measurement fidelities can in principle be compensated by
normalizing the data, but this was not done here for clarity.

In conclusion, we have introduced a new design for superconducting
phase qubits that explicitly separates the capacitive from the
inductive element of the Josephson junction. The number of TLS that
couple to the qubit is reduced by an order of magnitude, improving
the measurement fidelity to $90 \%$ and enabling quantum state
tomography. We believe that all aspects of our qubit's performance
may now be sufficient to demonstrate violations of Bell's
inequalities in coupled qubits. Furthermore, our results pave a
clear path towards future improvements in decoherence through the
fabrication of simple capacitors with lower dielectric loss.

We acknowledge Steve Waltman and NIST for support in building the
microwave electronics. Devices were made at the UCSB and Cornell
Nanofabrication Facilities, a part of the NSF funded NNIN network.
N. K. acknowledges support of the Rothschild fellowship. This work
was supported by ARDA under ARO grant W911NF-04-1-0204 and NSF under
grant CCF-0507227.


\begin{thebibliography}{8}
\bibitem{tlsnote} Previous work \cite{Martinis05} showed that
coherence loss scales as $N_c^2$, where $N_c$ is the average number of
individual TLS that couple to the qubit. The quantity $N_c$ scales as
the junction area $A_J$ multiplied by the largest splitting size
$S_{\mathrm{max}} \propto 1/\sqrt C$, where $C$ is the total
capacitance. Loss in coherence thus scales as $A_J^2/C$.
\end{thebibliography}
\end{document}